# A NOTE ABOUT THE SIMULATION PROGRAMS FOR HEAT AND MOLECULAR PHYSICS LABORATORY

Gabriel Murariu[1], Dorina Toma[1]

[1] *Faculty of Science, University "Dunărea de Jos", Galați, România*

*gabriel_murariu@yahoo.com*

**Abstract**

In order to understand the physics phenomena on the fundamental aspects, the software simulations are a good exercise in succeed of this desire. Some works of heat transport and molecular physics are studied in a comparative way (experiment vs. software applications). The main objects structure and some program interfaces are presented. Some screen – shot pictures are included.

PACS {01.30.Pp. 01.40.Gm, 01.50.Lc}

## 1. Introduction

In the past few years, many physical phenomena are studied in a new way: using computer simulation programs. This strategy means a very powerful tool in understanding the concepts and relations between physical systems parts. Other advantage is given by the low-cost for using this kind of physics study. However, this way implies a huge work volume to realize and simulate the whole set of studied system proprieties. In our present, teams of physicists and computer programmers raise this work. For our set of simulation programs for the heat transport and molecular physics works, it was necessary some years of work. In this paper is presented some results in the aim of succeeding in realize some applications of computational and interactive molecular physics. The last generation of built objects uses the C++ code and contains twelve different kind of interacting object. The first one use a Pascal code built objects and after that, we used a C++ compiler.

For a thermodynamics system without mechanical interactions, the laws for calorimetric phenomena, the Newton's heat transfer law and the Fourier law are sufficient for these applications.

## 2. Method and samples

We started with a simple problem of heat transport in a solid medium. The aim was to realize an algorithm to succeed in simulate the time evolution of the temperature in any point of one solid infinite body coupled to a thermostat and immersed in a exterior fluid with constant temperature. The experimental results and the software computations should be the same. The first result was an application which can describe the stationary final state of this system [Fig. 1].

The idea was to realize a set of object-oriented programs in order to presents o virtual evolution of the studied phenomena, and if is possible, to offer a referential point on the experimental measured values.

First of all, the project started with a huge list of objectives for each laboratory experimental work. After that, the next level was to build a complex structure of object proprieties groups to succeed in this scope. In our vision, this stage needs the main work part. It had to obtain a complete structure for all of these punctual objectives.

Finally was selected a serial of twelve object proprieties groups (in a software point of view) for the entire set of experimental works groups. The next stage was to built up these parts and further, to raise the simulation programs for each experimental work.

First generation of these applications where written using a very old Pascal compiler. After that, because the complexity of used object was greater and some graphic interface was required, we used a C++ compiler.

The third generation of these applications required a complete work of design, in order to allow studying the main proprieties of gas, liquid and solid part in interaction. The results of this new design consist in a new set of interacting applications [Fig. 1-4].

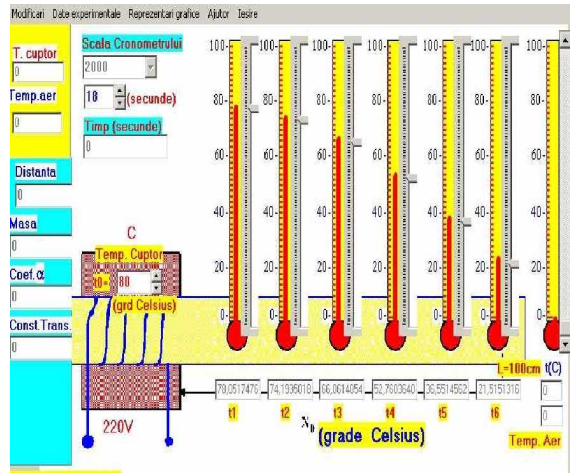 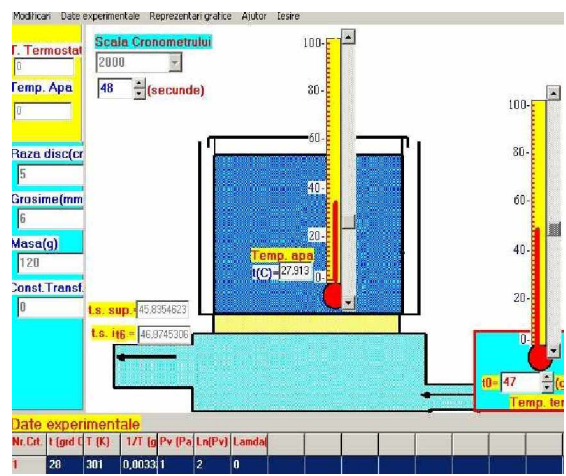

**Fig. 1**                                                                 **Fig. 2**

Can be simulated the reaching the stationary state for a metallic cylinder coupled with a thermostat [Fig.1] or the heat transfer from a solid body in order to find out the heat transfer coefficient (using the Newton's law) [Fig.2]. The temperature answer of a thermo-resistance [Fig.3] or the constant pressure caloric capacity of the real gases using heat transfer [Fig.4] can be presented in an interactive mod.

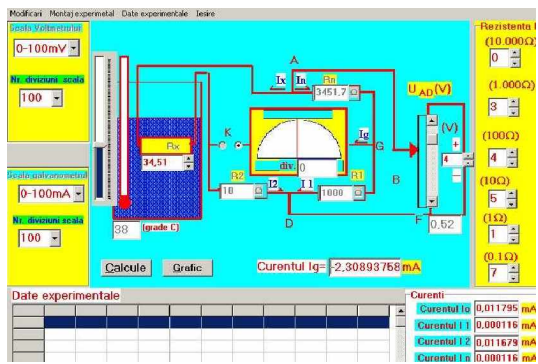 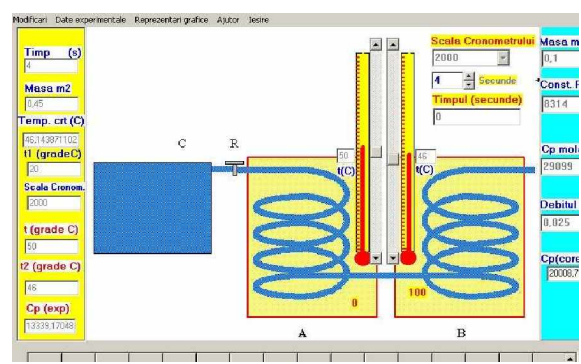

**Fig. 3**                                                                 **Fig. 4**

The model of this kind of process of only thermo-interacting parts is based on the scheme described in [Fig. 5].

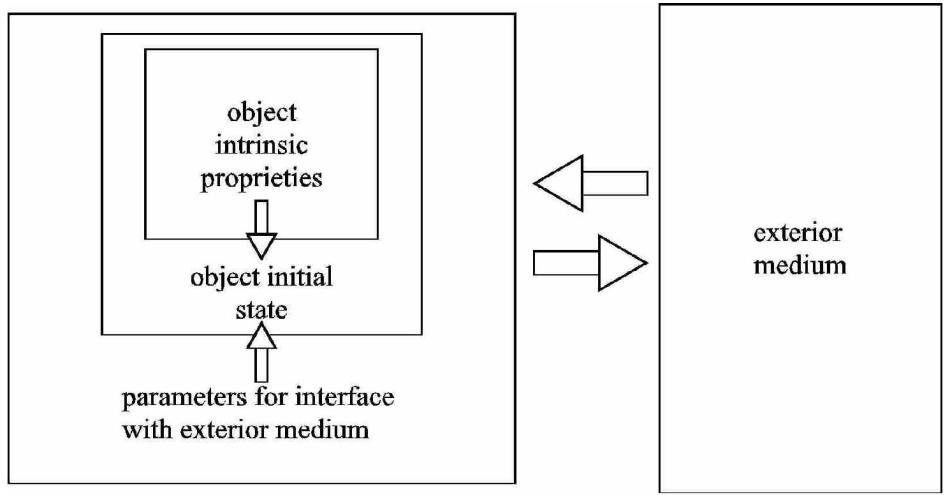

**Fig. 5**

In order to include the mechanical interactions it was developed the fourth applications generation, using a C++ compiler and graphical interface. Using the developed software objects can be studied for example, the ideal and real gases transformations [Fig. 7] or atmospheric humidity level [Fig.6].

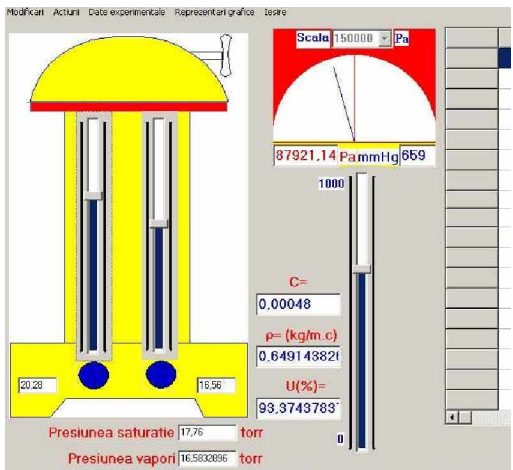
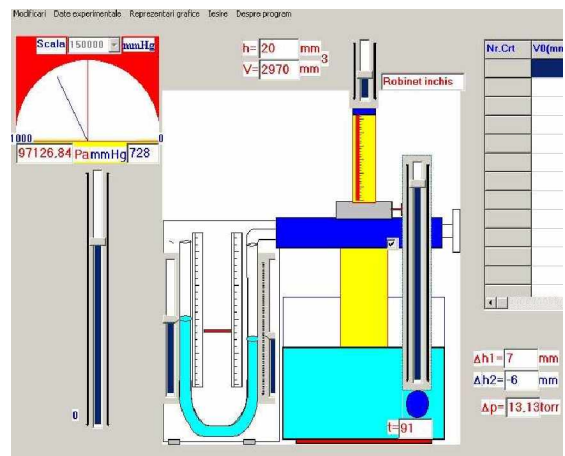

**Fig. 6**  **Fig. 7**

For these general thermodynamic interactions, the scheme is described in [Fig. 8]. The main part (object intrinsic proprieties) includes the equations of state.

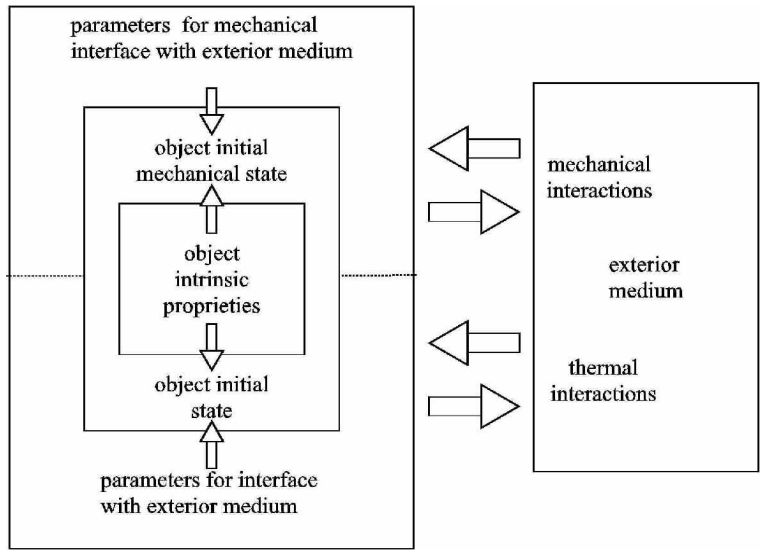

**Fig. 8**

The first level of the structure is the object which contains the intrinsic proprieties of the system part (like mass, density, etc.).

The next level contains the set of state parameters (as a double list – initial and final state parameters). At this level, it can be introduced proprieties like temperature, pressure, etc.

These are computed at each time using the information from the third level – the interface with the exterior medium. This level contains the information for the heat transfer coefficient, exterior aria, etc. One simple thermodynamic system is compounded from such parts, with a list of interactions between them. The algorithm must compute, starting from the initial state, using the intrinsic proprieties of each part, the time evolution of the considered system, toward the equilibrium final state. For example, now can be computed the final temperature for a system with an initial general temperature distribution, coupled with a medium with a constant temperature. This can be useful in the case of computation the transfer coefficients between different system parts.

For a greater interest some schematic interfaces were designed for every such application.

### 3. Conclusions

This result made us to keep trying in new areas. Every new idea which can make us to start a project in this way is welcome.

If this subject is interesting, it will be discussed and presented in an extensive view, with some software listings.

**Acknowledgements**

The author wishes to thanks to Sorin Brici for help and fruitful discussions. Special thanks go to Gh. Puşcaşu for his important design work and for his help to succeed in finalizing all the twenty-five thermodynamics applications.

**References**

[1] D. Toma, G. Murariu, *Fizică moleculară şi căldură*, Ed. Fundaţiei *"Dunărea de Jos", Galaţi*, (2002)

[2] D. Toma, G. Murariu, *Fizică moleculară şi căldură – Lucrări de laborator*, Ed. Fundaţiei *"Dunărea de Jos", Galaţi*, (2003)

[3] G. Murariu, D. Toma, *The XX-th Educational Conference*, Iaşi, (2002)

[4] G. Murariu, *The 6-th International Conference on Technical Informatics,* Timişoara, (2004)

[5] G. Murariu, N. Tigau, "On the simulation physics laboratory" *"Analele Universitatii de Vest",* Timişoara, (127 – 130), (2004)